\documentclass[a4paper,twocolumn]{esapub} 
\usepackage{natbib}
\usepackage{graphicx}

\title{Enhanced Pitch Angle Diffusion due to Electron-Whistler 
Interactions during Disturbed Times}
\author{W. Wykes}
\author{S. C. Chapman}
\author{G. Rowlands}
\affil{Space and Astrophysics Group, University of Warwick, Coventry, UK}

\newcommand{\vperp}{v_{\perp}}
\newcommand{\vpar}{v_{\parallel}}

\newcommand{\etal}{{\em et al.}}

\begin{document}

\maketitle

\begin{abstract}
During periods of increased magnetospheric activity, whistler wave emissions
have been observed with increased wave amplitudes. We consider a pitch angle
diffusion mechanism that is shown to scale with whistler wave amplitude and 
hence is 'switched on' during these periods of intense activity.

We consider the interaction between relativistic electrons and two oppositely 
directed, parallel propagating whistler mode waves. We show that for intense 
whistlers this process is stochastic and results in strong electron pitch 
angle diffusion.

We show that the interaction is rapid, occur on timescales of the order of 
tens of electron gyroperiods and that the interaction is sensitive to wave 
amplitude, wave frequency and electron energy.

\end{abstract}

\section{Introduction}

Electron whistler interactions have long been considered as a mechanism
for pitch angle scattering. Gyroresonance processes with near-parallel 
propagating waves (e.g. \citet{kennel}, \citet{lyons:84}) have been shown to 
produce pitch angle diffusion for electrons that are at resonance with a 
series of whistler waves (see \citet{gendrin}). We generalize resonant 
diffusion to include all phase space dynamics, i.e. as well as considering 
the resonant diffusion of trapped electrons we consider diffusion of 
untrapped electrons (we refer to this process as `off-resonance' diffusion). 
Therefore we maximize the area in phase space contributing to pitch angle 
diffusion.

The underlying dynamics of the interaction between electrons and a single 
whistler are inherently simple, as demonstrated by \citet{laird} who derived 
the Hamiltonian for relativistic electrons interacting with a whistler with a 
single wave number. However, for a single monochromatic whistler the process 
is not stochastic. We introduce stochasticity by including an identical, 
oppositely directed whistler mode wave.

We initially consider a simplified system consisting of monochromatic 
whistlers in order to understand the underlying behaviour. This treatment is 
then extended to consider whistler wave packets, i.e. a group of waves with a 
range of frequencies.

We derive approximate equations in the limit of low whistler wave amplitudes 
and consider the degree of pitch angle diffusion for waves of different 
frequencies and bandwidths and for electrons of different energies.

\section{Equations of Motion}

We derive full relativistic equations of motion and approximate them in the 
limit of low wave amplitudes, for the case of monochromatic whistlers and 
wave packets. We normalize time to the electron gyrofrequency 
$\Omega_e=eB_0/m_e$, (where $B_0$ is the background magnetic field), the 
wave amplitude is normalized to the background magnetic field, wave frequency
is normalized to the gyrofrequency and we normalize the electron velocity to 
the phase velocity of the waves, given by the electron dispersion relation 
(ignoring ion effects):
\
\begin{eqnarray}
\frac{k^2c^2}{\omega^2}=1-\frac{\omega_{pe}^2}{\omega(\omega-\Omega_e)} 
\label{eq:disp}
\end{eqnarray}
\
where $\omega_{pe}$ is the plasma oscillation frequency. Electrons can become 
trapped by either of the two whistlers if they are traveling close to the 
resonance velocity, given by:
\
\begin{eqnarray}
\omega-{\bf k} \cdot {\bf v_r}=n \Omega_e/\gamma \label{eq:res}
\end{eqnarray}
\
where $n$ is an integer and $\gamma = (1-v^2/c^2)^{-1/2}$ is the relativistic 
factor.

\begin{figure*}[phtb]
\centering
\includegraphics[width=1.0\linewidth]{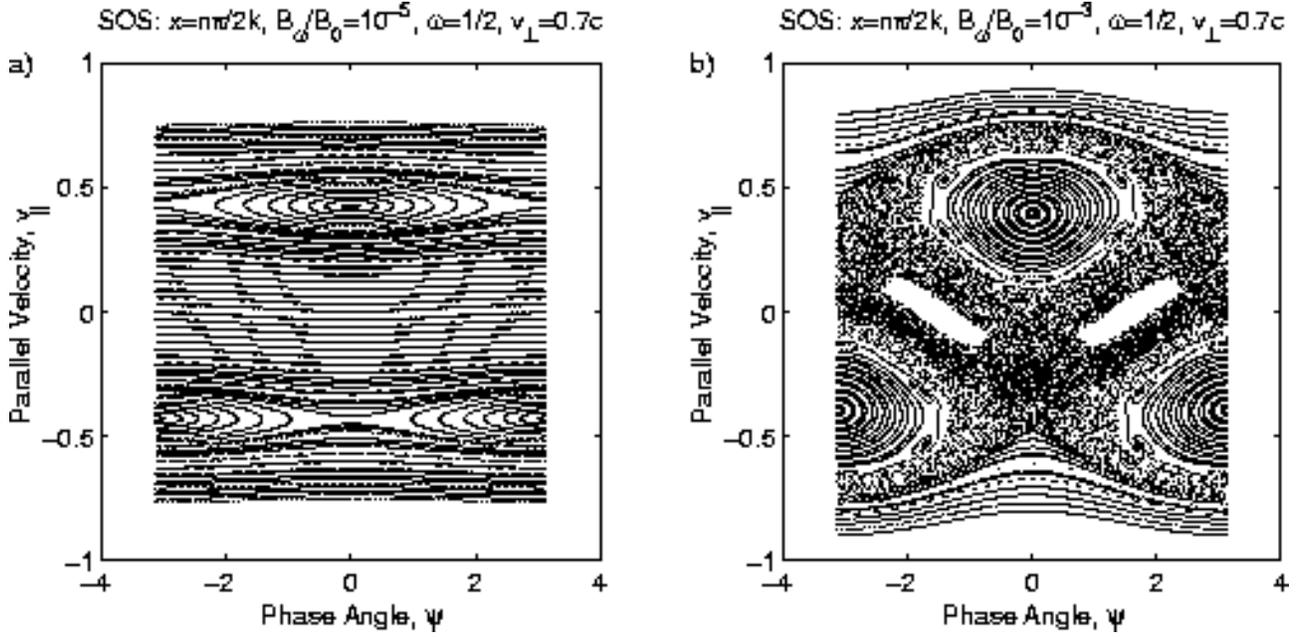}
\caption{\label{fig:mono} Stroboscopic surface of section plots for the 
monochromatic whistler interaction. For quiet time wave amplitudes, panel a), 
all trajectories are regular, with resonances given by the resonance condition
Equation~(\ref{eq:res}). For high wave amplitudes observed during disturbed 
times, panels b), phase space is dominated by stochastic trajectories with 
regular trajectories confined to close to the resonances. The stochastic 
region is bounded above and below by the first regular, untrapped, 
trajectories away from resonance, therefore there is a limit on the diffusion 
of electrons in phase space.}
\end{figure*}

\subsection{Monochromatic Whistlers}

In the limit of low wave amplitudes the full equations of motion can be 
reduced to:
\
\begin{eqnarray}
\frac{d^2 x}{d t^2} &=& \frac{2b {\vperp}_0}{\gamma_0}
\sin\left[(1/\gamma_0-\omega)t\right] 
\cos\left[kx\right]\label{eq:red}
\end{eqnarray}
\
where $x$ is the distance along the background magnetic field, $b$ is the
normalized wave amplitude, ${\vperp}_0$ is the initial perpendicular velocity 
and $\gamma_0=1/\sqrt{(1-{\vperp}_0^2/c^2)}$.

\subsection{Wave Packet Approximation}

Instead of a pair of waves it is more realistic to consider the interaction 
of a wave packet, ie a small group of waves with a range of wave frequencies
and wave numbers. We assume the wave amplitude is non-zero over wave 
frequency range $\Delta\omega$:
\
\begin{eqnarray}
b(\omega)=\left\{ \begin{array}{r@{\quad:\quad}l}
     0 & \omega<\omega_0-\Delta\omega/2 \\
     b & \omega_0-\Delta\omega/2<\omega<\omega_0+\Delta\omega/2 \\
     0 & \omega>\omega_0+\Delta\omega
     \end{array} \right. \label{eq:that}
\end{eqnarray}
\
where $\omega_0$ is the central wave frequency of the wave packet. Integrating
the monochromatic ehistler equation~(\ref{eq:red}) over the frequency 
range $\Delta\omega$ gives the following wave packet equation:
\
\vspace{1cm}

\begin{eqnarray}
\!\!\!\!\!\!\!\!\frac{d^2 x}{d t^2} \!\!\!\!\! &=& \!\!\!\!\!
\Omega_e \!\!\frac{2bv_{\perp 0}}{\gamma_0}\!\!
\sin\!\!\left[\!(\!\frac{1}{\gamma_0}\!-\!\omega)t \!+\! k_0 x \!\right]\!\!
\frac{\sin\!\left[\!(t\!-\!\beta x)\!\Delta\omega\!/\!2\right]}
{(t-\beta x)} \nonumber \\
&+& \!\!\!\!\!
\Omega_e \!\!\frac{2bv_{\perp 0}}{\gamma_0}\!\!
\sin\!\!\left[\!(\!\frac{1}{\gamma_0}\!-\!\omega)t \!-\! k_0 x \!\right]\!\!
\frac{\sin\!\left[\!(t\!+\!\beta x)\!\Delta\omega\!/\!2\right]}
{(t+\beta x)} \label{eq:packet}
\end{eqnarray}
\
where $1/\beta=d\omega/dk$ is the group velocity of the waves. The wave packet
equation~(\ref{eq:packet}) yields the monochromatic whistler 
equation~(\ref{eq:red}) in the limit $\Delta\omega \rightarrow 0$, with 
amplitude $b^{\prime}=b\Omega_e\Delta\omega$. 

\section{Numerical Results}


The monochromatic and wave packet equations were solved numerically using a 
variable order, variable stepsize differential equation integrator. We 
consider physical parameters for the terrestrial magnetosphere at $L=6$: 
gyrofrequency, $\Omega_e=25.3~kHz$, plasma frequency, $\omega_{pe}=184~kHz$, 
background magnetic field, $B_0=144~nT$ and wave amplitude 
$B_{\omega}=0.5~pT$, giving a normalized wave amplitude, consistent with 
quiet times in the terrestrial magnetosphere, $b=10^{-5}$ (see for example 
\citet{nagano}, \citet{parrot} and \citet{summers}).

The phase plots in Figure~\ref{fig:mono} are comprised of stroboscopic 
surfaces of section \citep{bene} to sample the full electron phase space. The 
initial parallel velocity was varied over the range $[-v_r,v_r]$, where $v_r$ 
is the resonance velocity, given by the resonance condition, (\ref{eq:res}), 
for $n=1$. All electrons were given a constant initial perpendicular velocity,
with $\vperp \approx 20 v_r$ as it was found that a high velocity anisotropy 
was required for stochasticity. 

In Figure~\ref{fig:mono} we plot parallel velocity $\vpar=dx/dt$ against phase
angle $\psi$, where $\psi$ is the angle between whistler propagating in a 
positive direction along the background field and the perpendicular velocity,
$\vperp$. In panel a) we consider a whistler wave amplitude consistent with
quiet times $(b=10^{-5})$ all trajectoriesin phase space are regular. There 
is little change in $\vpar$ and hence only weak pitch angle diffusion. As the 
wave amplitude is increased, stochastic trajectories are introduced, as the 
regular trajectories between the two resonances are progressively broken down.
In panel b) we consider the case of intense whistler wave activity during
substorms ($b=10^{-3}$, see for example \citet{parrot} and \citet{nagano}).
The stochastic region grows to encompass the resonances as the wave amplitude 
is increased. Regular trajectories are confined to KAM surfaces 
(near-integrable trajectories with an approximate constant of the motion 
\citep{tabor}). The stochastic region is bounded by the first untrapped 
(regular) trajectories away from the resonances, thus there is a limit on 
diffusion in phase space. 

As well as resonant diffusion of trapped electrons, there is diffusion 
of untrapped electrons throughout the stochastic region of phase space. Since,
for sufficient wave amplitudes, the stochastic region can encompass the 
resonances, the diffusion of untrapped electrons, which we refer to as 
`off-resonance' diffusion, may be enhanced over resonant diffusion. In 
addition we achieve pitch angle diffusion from a larger area of phase 
space.

\begin{figure}[phtb]
\centering
\includegraphics[width=1.0\linewidth]{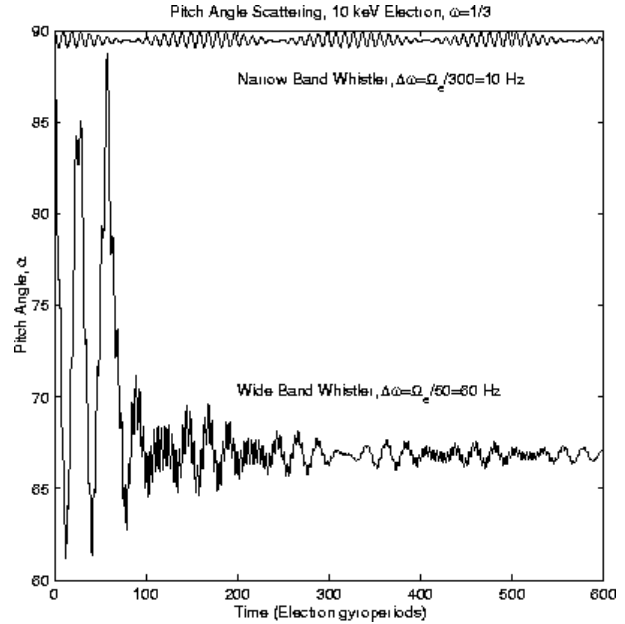}
\caption{\label{fig:nw} Change in pitch angle from an initial pitch angle of 
$90^{\circ}$, for quiet time wave amplitudes ($b=10^{-5}$) and narrow and 
wide whistler wave packets. For narrow wave packets there is little change in 
pitch angle. For wide wave packets there is a large change in pitch angle 
($\Delta\alpha\sim25^{\circ}$) occuring within a few tens of electron 
gyroperiods ($\sim10~ms$), hence the interaction is rapid. Changes in pitch 
angle attenuate with time and the pitch angle reaches a constant value.}
\end{figure}

Due to the time dependent nature of the wave packet equation~(\ref{eq:packet})
it is not possible to construct phase diagrams as in Figure~\ref{fig:mono} for
the monochromatic whistler case. Instead we can consider the dynamics of 
single electrons. In Figure~\ref{fig:nw} we show a single trajectory solution 
of the wave packet equation~(\ref{eq:packet}), for quiet time wave amplitudes 
and wide ($\Delta\omega=\Omega_e/50=500~Hz$) and narrow 
($\Delta\omega=\Omega_e/500=50~Hz$) whistler wave packets (see for example 
\citet{carpenter}).
We consider the change in pitch angle from an initial pitch angle of 
$90^{\circ}$. For narrow wave packets there is little change in pitch angle 
and the trajectory is regular. For wide wave packets the trajectory is 
stochastic with a large change in pitch angle ($\Delta\alpha\sim25^{\circ}$) 
occuring within a few tens of electron gyroperiods ($\sim10~ms$). We can now
achieve strong pitch angle diffusion for wave amplitudes consistent with the 
quiet time magnetosphere.

\begin{figure}[phtb]
\centering
\includegraphics[width=1.0\linewidth]{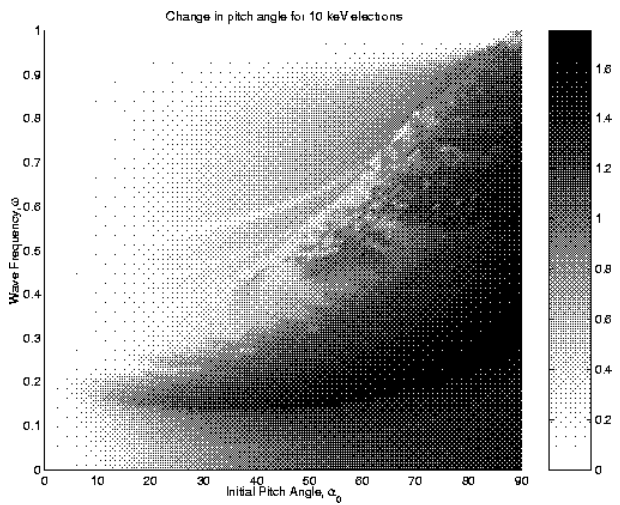}
\caption{\label{fig:10} Log change in pitch angle, 
$(\log_{10}|1+\Delta\alpha|)$, as a function of wave 
frequency, $\omega$, and initial pitch angle, $\alpha_0$,  for 10~keV 
electrons ($v= 0.2c$) and quiet time (low amplitude),wide band whistler wave 
packets ($\Delta\omega=\Omega_e/50$). For high to moderate initial pitch 
angles ($\alpha_0=50^{\circ}-90^{\circ}$) there is a change in pitch angle of 
up to $40^{\circ}$. For low pitch angles ($\alpha_0=5^{\circ}-10^{\circ}$) 
the change in pitch angle is of the order of a few degrees.}
\end{figure}

\section{Pitch Angle Scattering}

\begin{figure}[phtb]
\centering
\includegraphics[width=1.0\linewidth]{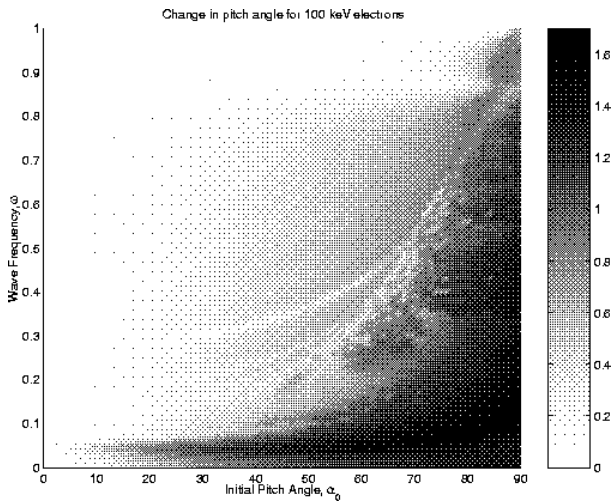}
\caption{\label{fig:100} As for Figure \ref{fig:10} except for 100~keV 
electrons ($v=0.5c$). Pitch angle scattering from a few degrees up to 
$40^{\circ}$ occurs although lower frequency whistler wave packets are 
required.}
\end{figure}

Using the wave packet equation~(\ref{eq:packet}) with quiet time whistlers
($b=10^{-5}$) and a relatively wide band whistler 
($\Delta\omega=\Omega_e/50=500~Hz$, \citep{carpenter}), we can estimate the 
degree of pitch angle 
scattering. In Figures \ref{fig:10} and \ref{fig:100} we estimate the  log 
change in pitch angle, $(\log_{10}|1+\Delta\alpha|)$, as a function of wave 
frequency, $\omega$, and initial pitch angle, $\alpha_0$. We consider the 
interaction between $10 keV$ electrons in Figure~\ref{fig:10} ($100 keV$ 
electrons in Figure~\ref{fig:100}) and wide band whistlers 
($\Delta\omega=\omega_e/50$). For high to moderate initial pitch angles 
($\alpha_0=50^{\circ}-90^{\circ}$) there is a change in pitch angle of up to 
$40^{\circ}$. For low pitch angles ($\alpha_0=5^{\circ}-10^{\circ}$) the 
change in pitch angle is of the order of a few degrees. In 
Figure~\ref{fig:100} we see a similar degree of diffusion except that lower 
frequency whistler wave packets are required.

\section{Discussion}

We have considered electron-whistler wave particle interactions to investigate
diffusion over all phase space, to include both resonant and `off-resonance'
diffusion. We have considered a simplified interaction with monochromatic 
whistler wave to understand the underlying behaviour and have shown that the 
presence of the second whistler wave introduces stochastic effects into the 
system. For wave amplitudes consistent with disturbed times we have shown 
that `off-resonance' diffusion occurs and that resonant diffusion is unchanged.

We have considered a more realistic case of whistler wave packets and have 
shown that for relatively wide band whistler wave packets strong pitch angle 
diffusion occurs for wave amplitudes consistent with quiet, undisturbed, 
times. For high initial pitch angles we estimate a change in pitch angle of up
to $40^{\circ}$, while for low pitch angles a change of a few degrees is 
estimated.

The effectiveness
in scattering electrons of different energies is dependent on the wave 
frequency. Electrons with low energies ($10~keV$) are readily scattered by 
waves of around half the electron gyrofrequency, while electrons at higher 
energies ($100~keV$) are scattered by lower frequency wave 
($\omega\sim\Omega_e/10$). $MeV$ electrons would require extremely low 
frequency waves for efficient scattering, hence our mechanism is most 
efficient for electrons in the $10-100~keV$ range.

\section*{Acknowledgements}
The authors would like to acknowledge PPARC for the funding of this work.

\end{document}